\documentclass[prb,twocolumn]{revtex4}   

\usepackage{amsmath}    
\usepackage{graphicx}   

\begin{document}
\title{Optical spectroscopy of plasmons and excitons in cuprate superconductors}
\author{D. van der Marel}
\affiliation{$^1$D\'epartement de Physique de la Mati\`ere Condens\'ee, Universit\'ee de Gen\`eve, CH-1211
Gen\`eve 4, Switzerland}
\date{\today}
\begin{abstract}
\pacs{} An introduction is given to collective modes in layered, high T$_c$ superconductors. An experimental
demonstration is treated of the mechanism proposed by Anderson whereby photons travelling inside the
superconductor become massive, when the U(1) gauge symmetry is broken in the superconductor to which the photons
are coupled. Using the Ferrell-Tinkham sumrule the photon mass is shown to have a simple relation to the
spectral weight of the condensate. Various forms of Josephson plasmons can exist in single-layer, and bi-layer
cuprates. In the bi-layer cuprates a transverse optical plasma mode can be observed as a peak in the $c$-axis
optical conductivity. This mode appears as a consequence of the existence of two different intrinsic Josephson
couplings between the CuO$_2$ layers. It is strongly related to a collective oscillation corresponding to small
fluctuations of the relative phases of the two condensates, which has been predicted in 1966 by A.J. Leggett for
superconductors with two bands of charge carriers. A description is given of optical data of the high T$_c$
cuprates demonstrating the presence of these and similar collective modes.
\end{abstract}
\maketitle
\section{Introduction}
Electrons form, together with the atomic nuclei, the basic fabric of materials. In order to expose the
organizing principles of matter, experimental physicists take apart the complicated fabric of matter, aimed with
a vast array of different spectroscopic methods. Spectroscopic tools typically expose the sample to an external
field or a beam of particles, and one measures the response of the sample to this external stimulus. Most of the
spectroscopic tools, such as optical absorption or inelastic scattering, do not reveal the nuclei or the
electrons directly. Instead one observes a spectrum of excited states which typically involve the excitation of
several or many electrons and/or nuclei simultaneously.

The reason is, of course, that the elementary particles forming a solid behave in a correlated way, and this is
already the case for the ground state of the material. As a result one can not excite a single electron without
influencing the state of the other particles in it's vicinity. Usually, if the amplitudes are not too large, the
excitations can be treated in the harmonic approximation. Regardless of the details of the material and of the
type of interactions between the particles one can, in principle and at least for small amplitudes, identify a
set of fundamental modes in the harmonic approximation. These so-called collective modes form an orthogonal set
of eigenstates of the material. To treat the electrical transport properties of metals it is usually much
simpler to refer to the language of electrons and holes. Nevertheless, even for simple metals like aluminum or
sodium, the metallic luster is caused by the plasma oscillations, which are one out of several possible
collective modes in a conducting material.

One of the relevant features of collective excitations is, that they provide the dynamical fluctuations
transforming between different states of matter. They can be populated either by varying the temperature or by
applying and external field, for example an electrical field, pressure, or magnetism. Broken symmetries are
typically accompanied by collective modes. We will now discuss a few examples:

(1) The phase of the order parameter in a superconductor is an example of a spontaneously broken U(1) symmetry.
This implies that the ground state is not unique but has a continuous degeneracy. In neutral superfluid the
fluctuations of this phase then possess linear dispersion\cite{bogoliubov59,nambu60,goldstone61}.

\begin{figure}[ht!]
\centerline{\includegraphics[width=7cm,clip=true]{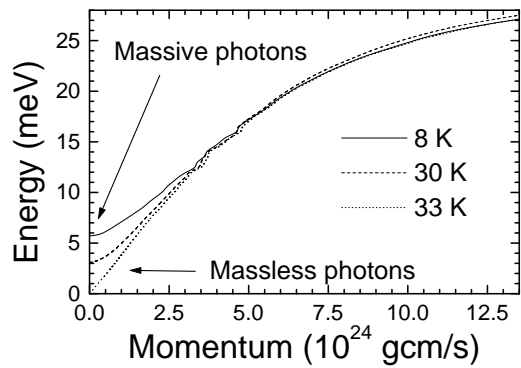}}
\caption{Energy-momentum dispersion of photons polarized along the
c-direction in La$_{1.85}$Sr$_{0.15}$CuO$_{4+\delta}$ for
different temperatures. T$_c$ of this sample is 33 K. The photons
travelling inside the superconductor become massive, when the U(1)
gauge symmetry is broken in the superconductor to which the
photons are coupled. } \label{figandersonhiggs} \end{figure}

(2) Earlier Anderson had shown from the gauge-invariant treatment
required for the Meissner effect, and taking into account the long
range nature of the Coulomb interactions, that in a superconductor
the longitudinal modes are
massive\cite{pwa58meissner,pwa58rpashort,pwa58rpalong}, and that
the transverse electromagnetic waves travelling in a
superconductor acquire a mass due to their coupling to the
superconducting condensate. An experimental example of this effect
is shown in Fig. \ref{figandersonhiggs}. A detailed discussion of
these data follows in section \ref{sectionandersonhiggs}.

(3) Anderson's mechanism was later used in the context of
elementary particle physics to predict, among other things, the
occurence of a novel massive elementary particle due to
spontaneous symmetry breaking, the Higgs boson, and to show that
that the W and Z bosons acquire a finite mass due to the coupling
to the symmetry broken
Higgs-field\cite{pwa58rpashort,pwa58rpalong,pwa63mass,higgs64mass,higgs67mass}.
The analogy between the theory of superconductivity and the
electro-weak theory is summarized in table
\ref{tableandersonhiggs}.
\begin{table}[t!]
\centerline{\includegraphics[width=8cm,clip=true]{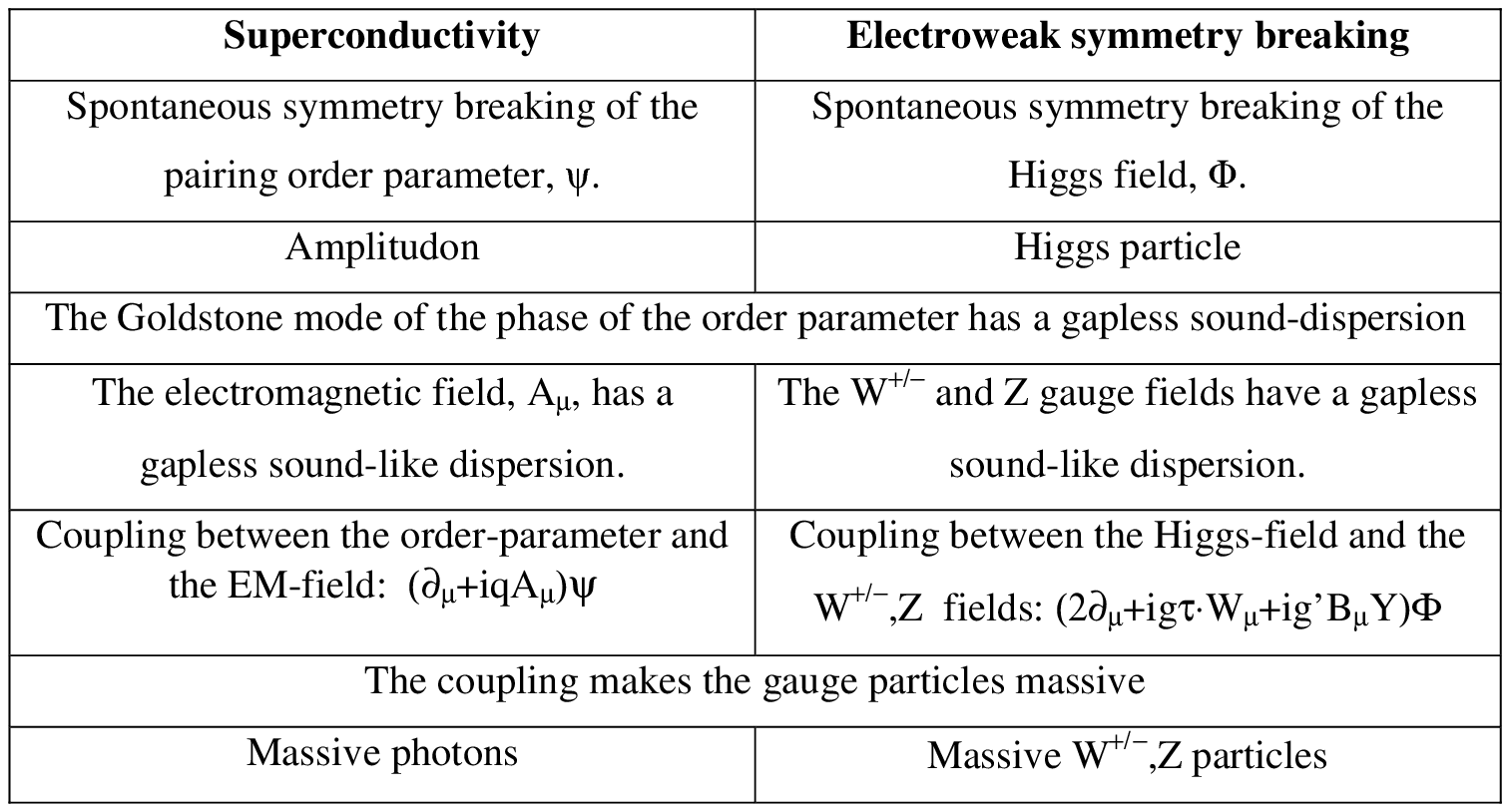}}
\caption{Some analogies between the theory of superconductivity
and the electro-weak theory.} \label{tableandersonhiggs}
\end{table}
\begin{figure}[b!]
\centerline{\includegraphics[width=7cm,clip=true]{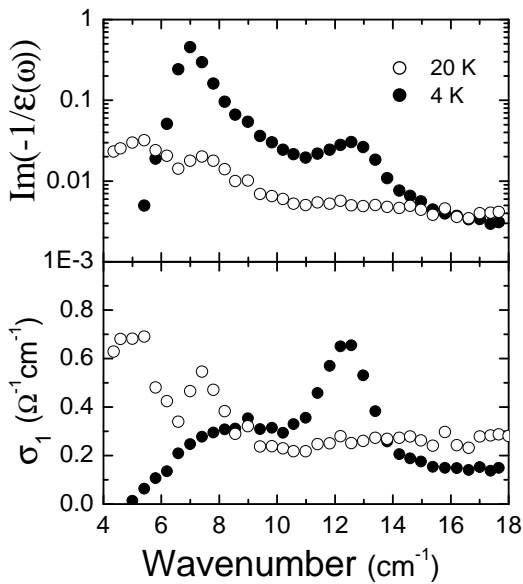}} \caption{The $c$-axis optical conductivity and
loss-function, of SmLa$_{0.8}$Sr$_{0.2}$CuO$_{4-\delta}$ for 4 K (closed symbols), and 20 K (open symbols).
T$_c$ of this sample is 16 K. When the sample enters the superconducting state, two longitudinal collective
modes appear (7 and 12.8 cm$^{-1}$) and one with transverse polarization (12.1 cm$^{-1}$). The two modes near 12
cm$^{-1}$ correspond to relative phase fluctuations of the two copper-oxygen layers within the unit
cell\cite{dulic01}.} \label{fig:slsco}
\end{figure}
\begin{figure}[h!]
\centerline{\includegraphics[width=7cm,clip=true]{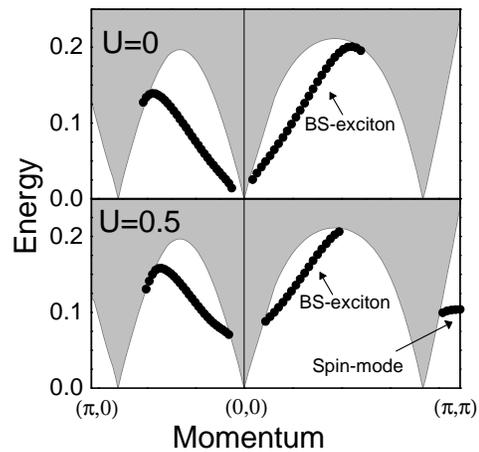}} \caption{RPA calculation of the collective modes of
a layered d-wave superconductor using a tight-binding calculation, reproduced from Ref.\onlinecite{marel95}.
Below the particle-hole continuum two types of modes occur: A fluctuation between d-wave and s-wave pairing
symmetry of the variety predicted by Bardasis and Schrieffer\cite{bardasis61} and a spin-fluctuation near the
$(\pi,\pi)$ point. The plasma-mode along the planes is at a much higher energy, not visible in this diagram.}
\label{eom}
\end{figure}

(4) The collective modes spectrum of the amplitude of the order
parameter of a superconductor has a gap, which has been observed
experimentally in NbSe$_2$ with Raman spectroscopy, and which
plays role equivalent to the Higgs particle in the electro-weak
theory\cite{nbse,varma1,varma2}.

(5) An unusual type of exciton has been predicted by Leggett for
the case were a superconducting gap occurs in two or more
overlapping bands\cite{leggett66} provided that a weak
Josephson-coupling between those bands is present. A similar type
of exciton is expected for the case where the crystal structure
contains pairs of weakly coupled two-dimensional
layers\cite{marel96prague,marel01}. For the latter case the
Coulomb interaction between the layers plays a dominant role. An
experimental example is shown in Fig. \ref{fig:slsco}. A detailed
discussion of these data follows in section \ref{slsco}.

(6) Motivated by the observation of a precursor infrared
absorption in Pb and Hg by Ginsberg, Richards, and
Tinkham\cite{ginsberg59}, Bardasis and Schrieffer have predicted
excitons in the superconducting gap, corresponding to pairing
symmetries different from those of the ground
state\cite{bardasis61,vaks62}. Fig. \ref{eom} shows an example of
such a mode in a model where the pairing-interaction has both an
s-wave and a d-wave channel. In the absence of a local repulsive
potential, the calculation predicts a soft excitonic mode near
$k=(0,0)$, which corresponds to a transition from s-wave to d-wave
order parameter. Increasing the on-site interaction results in an
increase of the energy of this exciton, implying that the d-wave
order parameter becomes more stable compared to s-wave symmetry.

(7) A paramagnetic Fermi-liquid is composed of two degenerate
liquids of opposite spin. The plasma-oscillations discussed above
correspond to an in-phase modulation of the two spin-densities.
The out-of-phase modulation is called a paramagnon. Because the
two spin-liquids oscillate out of phase, there is no net
charge-displacement, and consequently there is no restoring
electric force in contrast to the in-phase plasma oscillations.
These modes are therefore inside the particle-hole continuum and
they are normally overdamped. However, if the particle-hole
continuum is gapped, as happens in the superconducting state, a
paramagnon branch can occur below the particle-hole continuum,
with a correspondingly weak damping. In order to exist on an
energy scale below the superconducting gap, the paramagnon must be
very soft, implying that the system has been tuned close to a
spin-density wave instability. In Fig. \ref{eom} an example of
this fine-tuning is given, which was calculated using the
generalized random phase approximation scheme by
Anderson\cite{pwa58rpalong}, Bogoliubov, Tolmachev and
Shirlov\cite{bogoliubov59}. Increasing a local repulsive
interaction vertex $U$ from zero to 0.5, results in the emergence
of a soft spin-density mode near the $(\pi,\pi)$ point. Similar
behaviour has been observed with inelastic neutron scattering in
the cuprate superconductors\cite{bourges00a,bourges00b}, where
indeed a transition takes place to a spin ordered state when the
Mott-insulating state is approached by tuning the carrier
concentration.

(8) Various additional collective modes have been identified,
which are associated with a rotation between different order
parameters permitted by models containing additional symmetries.
Examples are the SO(4) symmetry of the negative U Hubbard
model\cite{demler96}, the SO(5) symmetry\cite{demler95} and the
SU(2) symmetry groups\cite{lee03}, where the latter two have been
proposed in the context of the t-J model. The complex order
parameter permitted by these models corresponds to a rich and
complicated spectrum of collective modes. The SO(5) model
generates a bosonic excitation with spin quantum number $S=1$ and
momentum $(\pi,\pi)$. This mode has been proposed for the
resonance with the same quantum numbers in the cuprates, which has
been observed with inelastic neutron scattering\cite{fong95}.

The superfluid phases of He$^3$ provide particularly beautiful
examples where several of the collective modes mentioned above
(and several others which are not in this list) have been observed
experimentally\cite{vollhardt}. In this article we will
concentrate mostly on collective modes which can be observed with
optical spectroscopy, i.e. items 1 to 6 of the previous list
involving flow of charge and current. Because the cuprates are
strongly correlated materials, and many of their properties can
not be explained within the context of the random phase
approximation, a large part of the subsequent discussion will be
based on classical field theory. The penalty one pays for this, is
that the properties which one can address with such a formalism
are limited to a particular set of collective modes. The advantage
is, that the results calculated with such a model do not heavily
rely on details on the microscopic level.

\section{Sound and plasmons}\label{soundandplasmons}

We begin by discussing the collective mode spectrum of a classical
compressible fluid of interacting particles of charge $e^*$ in a
charge-compensating background. The compressibility of the fluid
is $\kappa=n^{-2}\partial n/\partial \mu$, which is a scalar. The
mass of the particles, $m$, is in some cases an anisotropic
tensor. The fluctuations of the particle density around it's
equilibrium value are described by the field
$n(r,t)=n^{tot}(r,t)-n_0$. We furthermore allow the coupling of
the fluid to an electromagnetic field,
$\vec{E}(r,t)=-\vec{\nabla}\phi(r,t)-c^{-1}d\vec{A}(r,t)/dt,
\vec{B}(r,t)=\vec{\nabla}\times\vec{A}(r,t)$. The dynamical
behaviour of such a fluid can be described with the
Hamiltonian\cite{goldstein} \begin{eqnarray}\label{lagrangian} H=&
\int \left(\vec{\pi}-\frac{e^*\vec{A}}{c}\right)\cdot
\frac{n_0}{2\bar{m}}\cdot
\left(\vec{\pi}-\frac{e^*\vec{A}}{c}\right)d^3r +\int
\frac{|n(r)|^2}{2\kappa n_0^2}d^3r \nonumber \\
+& \int \int \frac{e^{*2}n(r)n(r')}{2|r-r'|} d^3r d^3r'
+ \int  e^*\phi(r)n(r) d^3r \\
\mbox{where}& \vec{\pi}\equiv\vec{\nabla}\nu(r)+\vec{\nabla}\times\vec{\mu}(r) \nonumber
\end{eqnarray}
The first set of Hamiltonian equations of motion for the longitudinal currents are $dn/dt=\delta {\cal H} /
\delta \nu =\partial {\cal H} /
\partial \nu-\vec{\nabla} \cdot \partial {\cal H} / \partial (\vec{\nabla}\nu)$, and $-d\nu/dt=\delta {\cal H}
/\delta n$. The second set for the transverse currents is $d\vec{\eta}/dt=\delta {\cal H} / \delta \vec{\mu}
=\partial {\cal H} / \partial \vec{\mu}+\vec{\nabla} \times
\partial {\cal H} / \partial (\vec{\nabla}\times \vec{\mu})$, and $-d\vec{\mu}/dt=\delta {\cal H} /\delta \vec{\eta}$.
The potential energy does not depend on $\vec{\eta}$, because a liquid has zero shear-modulus. This provides
four coupled relations between the currents and the density fluctuations
\begin{eqnarray}\label{lagrangianEOM}
-\frac{d}{dt}\nu(r)&=& \frac{n(r)}{\kappa n_0^2} + e^{*2}\int d^3r' \frac{n(r')}{|r-r'|} +e^*\phi(r) +
\frac{\nu(r)}{\tau}
\nonumber \\
-\frac{d}{dt}\vec{\mu}(r)&=&0
\\
\frac{d}{dt}n(r)&=&-\vec{\nabla}\cdot\frac{n_0}{\bar{m}}\cdot \left(
\vec{\nabla}\nu(r)+\vec{\nabla}\times\vec{\mu}(r)-\frac{e^*\vec{A}(r)}{c}\right)
\nonumber \\
\frac{d}{dt}\vec{\eta}(r)&=& \vec{\nabla}\times \frac{n_0}{\bar{m}} \cdot\left(
\vec{\nabla}\nu(r)+\vec{\nabla}\times\vec{\mu}(r)-\frac{e^*\vec{A}(r)}{c}\right) \nonumber
\end{eqnarray}
where the last term of the first line was introduced to represent the effect of dissipation on the current.
Combining these expressions we obtain the wave-equation
\begin{eqnarray}\label{sound}
n_0e^*\vec{\nabla}\cdot \frac{1}{\bar{m}}\cdot\vec{E}(r) +\left\{
\frac{d^2}{dt^2}+\frac{1}{\tau}\frac{d}{dt}\right\}n(r)
\nonumber\\
=\vec{\nabla}\cdot \frac{1}{\bar{m}}\cdot \vec{\nabla} \left\{ \frac{n(r)}{\kappa n_0} + e^{*2}n_0\int d^3r'
\frac{n(r')}{|r-r'|} \right\}
\end{eqnarray}
for the propagation of density fluctuations, or plasmons, of a charged fluid\cite{footnote1}. For plane waves
$\vec{E}(r,t)=\vec{E}_ke^{ik\cdot r-i\omega t}$, $n(r,t)=n_ke^{ik\cdot r-i\omega t}$, this amounts to
\begin{eqnarray}\label{sound-k}
\left( \omega^2+\frac{i\omega}{\tau} -
\vec{k}\cdot\left[\frac{\bar{\omega_p^2}}{|\vec{k}|^2}+\bar{v_s^2}\right]\cdot\vec{k} \right)n_k = i
\vec{k}\cdot\frac{e^{*}n_0}{\bar{m}}\cdot \vec{E}_k
\end{eqnarray}
where $\omega_p=(4\pi n_0 e^{*2}/m)^{1/2}$ is the plasma frequency and $v_s=(\kappa n_0 m)^{-1/2}$ is the
sound-velocity. We should keep in mind that $m$, $\omega_p^2$ and $v_s^2$ are tensors: when the mass-tensor is
anisotropic, the plasma-frequency and it's dispersion depend on the direction of propagation in the medium.

\section{Isotropic plasmon-dispersion}\label{isotropic}
Let us first consider the case of an isotropic charged fluid. In this case the plane wave solutions obey the
dispersion relation
\begin{equation}\label{sounddispersion}
\omega(k)^2=\omega_p^2+v_s^2k^2
\end{equation}
In a 3D Fermi gas the compressibility arises purely from the density of states at the Fermi energy:
$\kappa=n^{-2}dn/d\mu=3/(m v_F^2 n)$, and consequently the zero-sound velocity is\cite{mahan} $v_s=3^{-1/2}v_F$.
If we apply Eq. \ref{sounddispersion} to this case, we obtain $\omega_k = \omega_p+\frac{v_F^2}{6\omega_p}k^2 +
O(k^4)$ for the dispersion formula. Although this resembles the result obtained with the random phase
approximation (RPA) of the electron gas model\cite{mahan} $\omega^{RPA} = \omega_p+\frac{3v_F^2}{10\omega_p}k^2
+ O(k^4)$ the dispersive term in the RPA is a factor 9/5 larger, which shrinks to a value closer to 1 when
higher order electron-correlation diagrams are included in the calculation\cite{mahan,singwi68}. In the alkali
metals a systematic reduction has been observed with high energy electron energy loss spectroscopy as the
relative importance of the Coulomb interaction increases\cite{fink89}. It is important to point out here, that
at a qualitative level the dispersion of the plasma modes does not rely on the fact that microscopically the
particles in the fluid are fermions. Indeed, Eq. \ref{sounddispersion} expresses a rather generic feature of a
liquid, namely that it has a finite compressibility. For this reason Eq. \ref{sounddispersion} has a broad
applicability, which goes beyond the special case of a Fermi-gas. This becomes particularly important in cases
where on a microscopic level the properties are not fully understood, like in the cuprate materials: In spite of
the lack of a fully established microscopic framework it is still possible to predict a certain number of
properties at least at a qualitative level, in particular the collective plasma modes.

Superconductors are characterized by a macroscopic coherent state $\psi(r,t)$. Usually it is assumed that the
variations of the amplitude are negligibly small, hence $\psi(r,t)=n_0^{1/2}\exp{\{i\varphi(r,t)}\}$. In this
case the macroscopic current and the density of such a state are
\begin{subequations}\label{supercurrent}
\begin{eqnarray}
n(r,t)&=&|\psi(r,t)|^2 = n_0 \\
\vec{j}(r,t)&=&-\frac{n_0}{m}\hbar\vec{\nabla}\varphi(r,t)
\end{eqnarray}
\end{subequations}
The equations of motion and the dispersion relation of the plasma-modes in the superconducting state are also
given by Eqs. \ref{sound} and \ref{sounddispersion}, with the dissipation $1/\tau$ set to zero, but $v_s$ and
$\omega_p$ may differ from those in normal state.

\section{Strong anisotropy}\label{anisotropic}
Let us consider now the case of quasi-two dimensional materials, characterized by a large mass along
perpendicular to the planes and a light mass along it. For this case we obtain from Eq. \ref{sound-k} the
dispersion relation
\begin{eqnarray}\label{sounddispersionaniso1}
\omega(\vec{k})^2&=&\frac{\omega_{p,p}^2k_{p}^2+\omega_{p,s}^2k_{s}^2}{k_{p}^2+k_z^2}+v_{s,p}^2k_{p}^2
+v_{s,p}^2k_z^2
\end{eqnarray}
Inspection of this expression reveals, that $k=0$ corresponds to a singular point: If we approach it along the
planes, we obtain $\omega(k_p\rightarrow 0,k_z=0)= \omega_{p,p}$, while along the z-direction $\omega(k_p=
0,k_z\rightarrow0)= \omega_{p,z}$. In the extreme situation where the material is insulating along the
z-direction, one obtains
\begin{equation}\label{sounddispersionaniso2}
\omega(\vec{k})^2=\omega_{p,p}^2\frac{k_{p}^2}{k_{p}^2+k_z^2}+v_{s,p}^2k_{p}^2
\end{equation}
which corresponds to the small $k_z$ limit of the layered electron gas model\cite{fetter74,kresin88,bose92}. In
this case $\omega(k_p=0,k_z\neq 0) = 0$. Moreover, if we consider the dispersion along the plane, while keeping
$k_z$ fixed at a finite value, it becomes sound-like: $\omega=v_{eff}k_x$, with a sound velocity
$v_{eff}^2=\omega_{p,p}^2/k_z^2+v_{s,p}^2$. This behaviour can be measured with $k$-dependent electron energy
loss spectroscopy\cite{schulte00}.

\section{Dielectric function}\label{dielectric}
It is straightforward to solve Eq. \ref{sound} in the presence of a longitudinal external field
$\vec{E}^e(r,t)=-\vec{\nabla}\phi(r,t)$: The total electric field is the sum of the externally applied field and
the field arising from the charge distribution of the matter-field,
$\vec{\nabla}\cdot\vec{E}(r,t)=\vec{\nabla}\cdot\vec{E}^e(r,t)+4\pi e^*n(r,t)$. The longitudinal inverse
dielectric function, describing the response to an external charge, is defined as the external field divided by
the total field: $\epsilon=\vec{\nabla}\cdot\vec{E}^e/\vec{\nabla}\cdot\vec{E}$. With the help of Eq.
\ref{sound} the inverse dielectric function is now easily obtained
\begin{eqnarray}\label{charge response}
\frac{1}{\epsilon(\omega,k)}&=&1+\frac{4\pi e^*n(r,t)}{\vec{\nabla}\cdot\vec{E}^e(r,t)}
\nonumber\\
&=&1-\frac{\omega_p^2}{\omega_p^2+v_s^2k^2-\omega(\omega+i/\tau)}
\end{eqnarray}

The plasma-modes correspond to the condition that an arbitrarily weak density fluctuation with wave-vector $k$
and frequency $\omega$ can generate a finite electromagnetic response $\vec{E}$. Because $1/\epsilon(\omega,k)$
describes the response to the density fluctuation, in an isotropic fluid these modes have the electrical
polarization parallel to the propagation direction. The charge-density modes therefore correspond to the poles
of Eq. \ref{charge response}
\begin{equation}\label{modes}
\omega=\frac{1}{2i\tau}\pm\sqrt{\omega_p^2+v_s^2k^2-\frac{1}{4\tau^2}}
\end{equation}
The relevant limit for optical spectroscopy is $k\rightarrow 0$ for which the mode frequencies become purely
imaginary if $1/\tau>2\omega_p$. This corresponds to the limit of over-damping. This situation is characterized
by the fact that Re$\epsilon(\omega,k)>3/4$ for all (real) frequencies. On the other hand, the condensate of a
superconductor is characterized by the absence of dissipation, at least for frequencies much smaller than the
superconducting gap, implying that $\epsilon(\omega,k)=0$ {\em must} occur at some finite frequency. Hence in
materials where the dissipation in the normal state is large enough to cause an over-damped charge response, the
charge-density collective mode {\em must} emerge at some finite frequency within the frequency window of
dissipation-less flow when the material becomes superconducting at low temperatures.

\begin{figure}[ht]
\centerline{\includegraphics[width=7cm,clip=true]{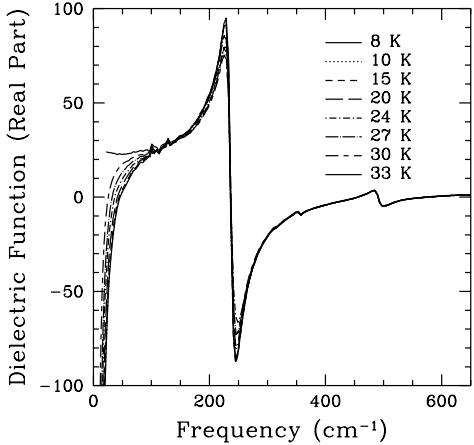}} \caption{Dielectric function with
polarization along the c-direction of La$_{1.86}$Sr$_{0.14}$CuO$_{4+\delta}$ for different temperatures. T$_c$
of this sample is 33 K. The data are from Ref. \onlinecite{kim95}.} \label{lsco_c_eps}
\end{figure}

This kind of behaviour is indeed observed for the c-axis optical response of the cuprate
superconductors\cite{tachiki94,fertig90,hwang95,bulaevskii95,tamasaku92,tsui94,matsuda95}. An example of this is
shown in Fig. \ref{lsco_c_eps}. T$_c$ of this crystal was 33 K. At this temperature $\epsilon'(\omega)$ never
crosses zero except for the phonons between 200 and 400 cm$^{-1}$. Below T$_c$ a dissipation-less low-frequency
electronic mode appears, characterized by a zero-crossing of $\epsilon'(\omega)$ reaching about 50 $cm^{-1}$ for
$T\ll T_c$. Note, that although superconductivity implies the presence of a plasma-mode, the reverse is not
true: In fact in most superconductors known to date, the plasmon is {\em not} overdamped in the normal state,
and the transition into the superconducting state affects the plasma-frequency only marginally.
\begin{figure}[ht]
\centerline{\includegraphics[width=7cm,clip=true]{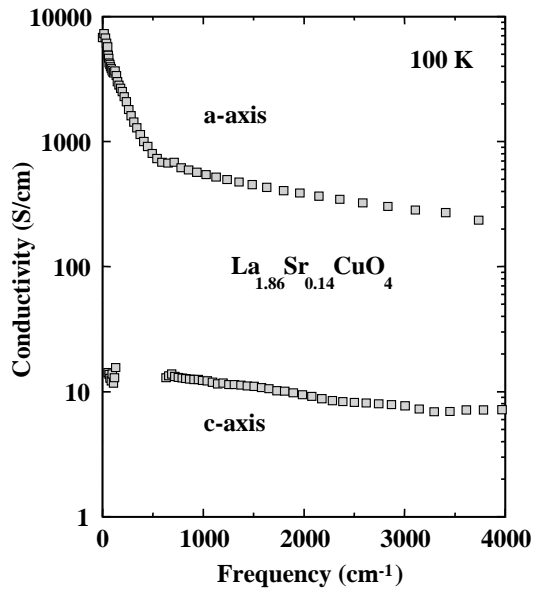}} \caption{Optical conductivity with
polarization along the planes and along the c-direction of La$_{1.85}$Sr$_{0.15}$CuO$_{4+\delta}$ for T=100 K.
To obtain a high accuracy for the optical conductivity along the c-direction, the transmission coefficient was
measured of a 20 $\mu$m thick slab, which had been cut along ac-plane. Due to a strong absorption by the optical
phonons between 200 and 600 cm$^{-1}$, no transmission could be detected in this range. Therefore the c-axis
optical conductivity is not shown in this range. The data are from Ref. \onlinecite{marel00}.}
\label{lsco_c&a_sigma}
\end{figure}

A second type of collective mode has the electric field perpendicular to the direction of propagation. Those
modes correspond to absorption peaks of the optical conductivity function, $\sigma(\omega)=j/E$, which in the
long wavelength limit is proportional to the dielectric function
$\sigma_1(\omega)=\omega\mbox{Im}\epsilon(\omega)/(4\pi)$. In the limit that $k\rightarrow 0$ this condition
requires, that an arbitrarily weak electric field results in a finite current. Hence at the transverse
collective mode frequency $\sigma(\omega)\rightarrow \infty$. For a single component plasma this happens at
$\omega=-i/\tau$. In a superconductor $1/\tau=0$, and the condensate causes indeed a diverging conductivity at
zero frequency. In a normal metal $\tau$ is finite, and the optical conductivity is characterized by a narrow
Drude peak. In the cuprates, if the electric field is polarized along the c-direction, a narrow Drude peak is
however {\em not} observed in the frequency dependence of $\sigma(\omega,T)$. Also the temperature dependence of
the c-axis conductivity is reminiscent of a semi-conductor, at least in samples which are not strongly
overdoped. Hence the cuprates have a strong anisotropy between ab-plane and c-axis conductivity in several
respects: (i) The DC-resistivity, (ii) the superfluid spectral weight of the superconducting state, (iii) the
temperature dependence of the conductivity (iv) and the frequency dependence of the conductivity. Although (i)
and (ii) can be easily explained from a large effective mass anisotropy, (iii) and (iv) imply that the transport
mechanism itself is anisotropic. These observations, which belong to the oldest and most firmly established
features of the cuprate superconductors, have been -and still are- the subject of many speculations, none of
which have been completely satisfactory in every respect. Fig. \ref{lsco_c&a_sigma} demonstrates an example of
this. The strong optical phonons obscure part of the electronic response along the c-axis, but it is clear from
this graph that the electronic response has only a weak frequency dependence, and does not appear to agree with
the Drude line-shape.

\section{The mass of a photon in a superconductor}\label{sectionandersonhiggs}
\subsection{The Anderson-mechanism for photon-mass
generation}\label{subsectionandersonhiggs}
For long wavelengths the energy momentum dispersion relation of transverse polarized electromagnetic waves
propagating inside the superconductor can be calculated from the relation\cite{footnote1} $k^2 c^2 =
\epsilon(\omega)\omega^2$ between the wave-vector, the frequency and the dielectric function. In Fig.
\ref{figandersonhiggs} the data of Fig. \ref{lsco_c_eps} have been displayed in this way. This is an example of
the Anderson-mechanism discussed in the introduction, which generates a finite mass of the photons by coupling
them to a spontaneous symmetry-breaking field. We can see from Fig. \ref{lsco_c_eps}, that compared to the W-
and Z-boson in elementary particle physics this is merely a feather: the photon-gap is about 6 meV, which is
$10^{13}$ times smaller than the mass of the W-boson (80 GeV). The estimated mass of the
Higgs-boson\cite{krawczyk98} is $10^{13}$ times larger than the superconducting gap, $\sim 2k_BT_c=5.5$ meV.

Using a cavity resonance technique operating at a single photon energy, and using a magnetic field to tune the
plasma-frequency of the superconductor inside the cavity, the presence of the mass-gap has been demonstrated
both for $k \parallel E
\parallel c$ and for $k\perp E
\parallel c$ in Bi2212\cite{kadowaki97,kadowaki98}.

\subsection{The Tinkham-Ferrell sumrule}
\begin{figure}[ht]
\centerline{\includegraphics[width=7cm,clip=true]{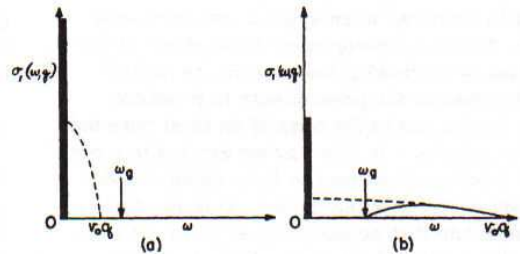}} \caption{Figure copied from Ref.
\onlinecite{tinkham59} by Tinkham and Ferrell. Original caption: "Effect of the superconducting transition on
the frequency-dependent conductivity for (a) long- and (b) short-wavelength transverse electromagnetic waves.
The normal-state conductivity is indicated by dashed curves and extends to the maximum frequency of $v_0q$,
where $v_0$ is the Fermi velocity and $q$ is the wave number. In (a) the wavelength is sufficiently long that
the maximum absorption frequency in the normal state falls short of the energy gap threshold $\hbar \omega_g$.
Consequently essentially all of the oscillator strength is absorbed by the delta function at zero frequency,
leading to a full London current. In (b) the shorter wavelength causes the absorptions in the normal state to be
spread over a frequency interval much larger than the energy gap. The strength of the delta function is
therefore less and the London current is weakened. This dependence of the London current on wavelength is
equivalent to the nonlocal current-field relation of Pippard." } \label{tinkham}
\end{figure}
A useful and important property of the c-axis optical conductivity concerns the spectral weight sum-rule, or
f-sum rule: (i) The spectral weight represented by the zero-frequency $\delta$-function in the superconducting
state is represented as the square of the plasma-frequency of the condensate, $\omega_{p,s}^2$. (ii) A
consequence of BCS theory is, that the spectral weight at finite frequencies is reduced due to the opening of a
gap in the optical conductivity. (iii) The Tinkham-Ferrel sumrule\cite{glover57a,glover57b,richards58,tinkham59}
illustrated in Fig. \ref{tinkham}, asserts that the spectral weight of the condensate balances exactly the
decrease of optical spectral weight integrated over all frequencies larger than zero, compared to the normal
state:
\begin{equation}\label{gtf}
\rho_{s} = \int_{0^+}^{\infty} \mbox{Re}\left(\sigma_n(\omega,\vec{q})-\sigma_s(\omega,\vec{q})\right) d\omega
\end{equation}
where the wave-vector $\vec{q}$ is introduced in a general framework for the analysis, where all space and
time-dependent quantities inside the material are Fourier-analyzed. The above stated sumrule has a general
validity, irrespective of the microscopic details, due to the fact that it follows from a strict conservation
law ({\em i.e.} particle number conservation). In recent years the sumrule has been an important instrument to
address ideas that superconductivity may be stabilized by interlayer tunneling, or, more generally, to study the
direction of change of kinetic energy when the material becomes superconducting both along the
c-direction\cite{chakravarty93,pwa95,ajl96,marel96houston,schuetzmann97,kam98,tsvetkov98,chakra98,
kirtley99,gaifullin99,basov99,chakravarty99,basov01a,munzar01,boris02,munzar03b}, and along the planar direction
\cite{hirsch92a,hirsch92b,molegraaf02,santander03,marel04a}.

\begin{figure}[ht]
\centerline{\includegraphics[width=7cm,clip=true]{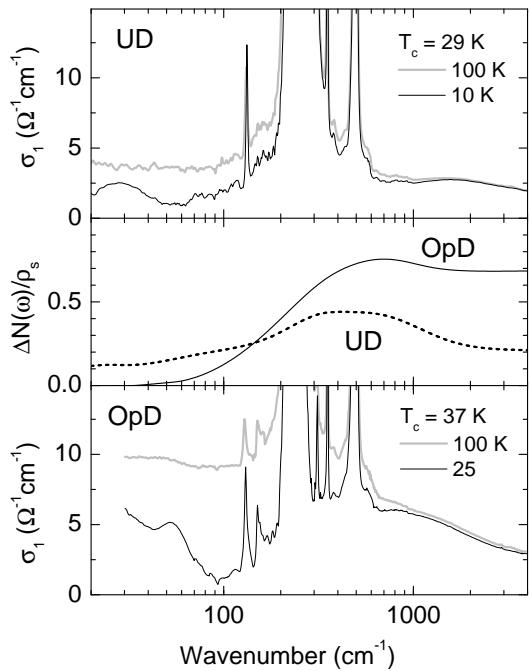}} \caption{Optical conductivity of underdoped (top
panel) and optimally doped LSCO (bottom panel). In the middle panel the sum-rule check is displayed. The
quantity displayed here corresponds to $(N^n(\omega,T)-N^{sc}(\omega,T))/\rho_{s}(T)$, where the temperature
dependent optical conductivity of the normal state has been extrapolated to obtain $N^n(\omega,T)$ and
$N^{sc}(\omega,T)$ both at the same temperature below $T_c$. The details of this analysis are given in Ref.
\onlinecite{kuzmenko03}.} \label{alexey}
\end{figure}

In a BCS superconductor one expects that by and large most of the spectral weight should be recovered on an
energy range of 4 times the gap\cite{footnoteBCS-sumrule}. Experiments with light polarized along the c-axis of
the cuprates have revealed that in the underdoped samples a large fraction of the spectral weight remains
unrecovered up to 20 times the gap energy\cite{basov99,basov01a,kuzmenko03}. At optimal doping the absolute
value of non-recovered c-axis spectral weight is even larger (see Fig. \ref{alexey}), but the ratio $\Delta
N(\omega_m)/\rho_{s}\equiv\rho_{s}^{-1}\int_{0^+}^{\omega_m}
\mbox{Re}\left(\sigma_n(\omega)-\sigma_s(\omega)\right) d\omega$ decreases due to the fact that $\rho_{s}$
increases sharply as the doping is increased. Interestingly, in addition to the opening of the superconducting
gap, we observed an {\it increase} of conductivity above the gap up to 270 meV with a maximal effect at about
120 meV\cite{kuzmenko03}. This may indicate a new collective mode at a surprisingly large energy
scale\cite{lee03}. For the $ab$-direction of underdoped and optimally doped Bi2212 about 0.25$\%$ of the
ab-plane spectral weight remains unrecovered up to about 20 times the gap energy\cite{molegraaf02}. However, in
absolute numbers the unrecovered spectral weight is orders of magnitude larger along the planes than along the
c-direction.

\subsection{Relation between photon-mass and spectral-weight}\label{sectionandersonhiggs2}
The presence of the superconducting condensate contributes a term $-8\rho_s/\omega^2$ to the dielectric
function. Because this term diverges for $\omega\rightarrow 0$, and $\epsilon(\omega)$ should become positive
for $\omega \rightarrow \infty$, this means that in the superconducting state the dielectric function {\em must}
cross through zero for some finite frequency. This is the plasma-frequency of the superconducting condensate. If
the remaining contributions to the dielectric function, $\epsilon_b$, (for example those coming from optical
phonons) have a weak frequency dependence in the region where this zero crossing occurs, the plasma frequency of
the condensate will be $\omega_{p,s}=(8\rho_s/\epsilon_b)^{1/2}$. In section \ref{subsectionandersonhiggs} we
have seen that, due to Anderson's mechanism for mass generation, inside a superconductor the photons acquire a
gap $\hbar \omega_{p,s}$. For the conditions described above the photon dispersion relation is
$\omega^2=(k^2c^2+8\rho_s)/\epsilon_b$. Because the dynamical mass of the photons is $m_{A} = \hbar/(d^2 \omega
/ dk^2)$, we conclude that the following relation exists between the spectral weight removed from the optical
conductivity implied by the Tinkham-Ferrell rule, and the photon-mass inside the superconductor discussed in the
previous section:
\begin{equation}
\rho_s=\frac{m_{A}^2c^4}{8\hbar^2\epsilon_b}
\end{equation}
In the example given in Fig. \ref{figandersonhiggs} we have $\rho_s \simeq 2.5 \cdot 10^{26}$ s$^{-1}$, and
$\epsilon_b\simeq 22.7$. Using the expressions above, the dynamical mass of c-polarized photons in this compound
is then 2.5$\cdot 10^{-34}$ g, which is about four-million times less than the mass of an electron.

\section{Multi-component plasma}

\begin{figure}[ht]
\centerline{\includegraphics[width=7cm,clip=true]{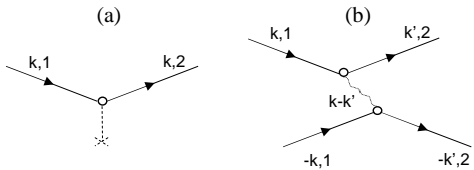}} \caption{(a) Interband coupling whereby a single
electron is transferred from band 1 to band 2, while conserving momentum. (b) Interband coupling process whereby
a pair of electrons is transferred from band 1 to band 2.} \label{feynman}
\end{figure}

The problem of a coupled two-component superconducting plasma has been studied by Leggett. In this model one
considers a homogeneous mixture of two liquids, representing two different bands of charge carriers ({\em e.g.}
an s-band and a d-band). Electrons can flow from one band to the other, which introduces a current between the
two bands, $\frac{d}{dt}(n_s-n_d)$. The coupling, indicated in Fig. \ref{feynman}b, involves the scattering of a
pair of electrons from the s-band to the d-band and {\em vice versa}. This is a particular (pairing) term in the
potential energy. Only if in addition the wave-functions differ on an atomic scale, the flow of charge from one
band to the other involves a real motion of electrons. In the somewhat analogous case of longitudinal NMR in
3-He-A, the "bands" are the up- and down-spin states, so it is clear that in that case there is no flow in real
space. This contributes a portion to the energy proportional to $(\frac{d}{dt}(n_s-n_d))^2$. The corresponding
Hamiltonionan is
\begin{eqnarray}\label{leggetthamiltonian}
\cal{H}&=&  \frac{\kappa n_{0}^2\Gamma^2}{4}
(\nu_{s}(r,t)-\nu_{d}(r,t))^2\nonumber\\
&+&\frac{1}{2\kappa n_{0}^2} (n_{s}(r,t)^2+n_{d}(r,t)^2)
\end{eqnarray}
where $\Gamma$ is a parameter which characterizes the coupling. The second term of the Hamiltonian is equivalent
to the term in Eq. \ref{lagrangian} describing the compressive strain energy of the two fluids, and can be
decomposed into a relative- and a total density fluctuation,
$2n_{s}^2+2n_{d}^2=(n_{s}-n_{d})^2+(n_{s}+n_{d})^2$, where only the former has a corresponding coupling term in
the Hamiltonian. With these definitions the Hamiltonian equations of motion are
\begin{subequations}\label{leggetthamiltonianEOM}
\begin{eqnarray}
\frac{d}{dt} (n_{s}-n_{s})   &=&   \kappa n_{0}^2\Gamma^2 (\nu_{s}-\nu_{d})
\\
\frac{d}{dt} (\nu_{s}-\nu_{d}) &=& - \frac{1}{\kappa n_0^2} (n_{s}-n_{s})
\end{eqnarray}
\end{subequations}
The solution has the form $\delta n(t)=\delta n(0)e^{i\Gamma t}$: This is a collective mode with a frequency
$\Gamma$, where charge oscillates between the two reservoirs.

Already in a normal metal with two partly occupied bands, excitations exist which correspond to the oscillation
of charge between the two bands: From the single electron band structure the lowest optical energy electron-hole
transition occurs at an energy corresponding to the smallest vertical distance between the two bands, for
k-values where there is one band above and one below the Fermi energy. This splitting is usually of the order of
the bandwidth, {\em i.e.} of the order of 1 eV, although in principle it can be zero provided that the two bands
accidentally cross at $E_F$ (this happens in some of the bucky-tubes).

\begin{figure}[ht]
\centerline{\includegraphics[width=7cm,clip=true]{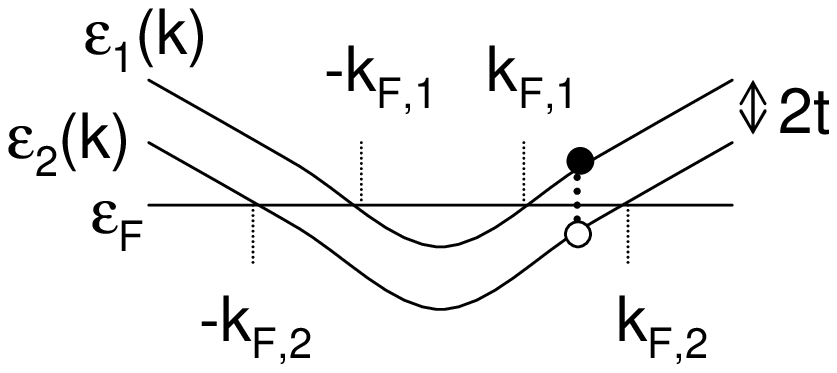}} \caption{A doubly degenerate band is split due
to a inter-band coupling term $t$. Electromagnetic radiation at a frequency $2t/\hbar$ can excite the electrons
across the split band.} \label{splitband}
\end{figure}

A special case occurs if the two bands are degenerate for all k-values: The presence of an interband-coupling,
$t$ scatters electrons between two different bands, while conserving their momentum, $k$. The eigenstates become
symmetric and anti-symmetric combinations of the two original bands at every k-point, split by an energy
difference $2t$ (see Figs. \ref{feynman}a and \ref{splitband}). In the optical spectra this splitting causes an
absorption band, corresponding to the 'vertical' ($\delta k=0$) excitations from the occupied 'symmetric' to the
unoccupied anti-symmetric band, as indicated in Fig. \ref{splitband}. This absorption is peaked at a frequency
$2t/\hbar$, which indicates that in this case $\Gamma$ represents the single particle inter-band coupling:
$\hbar\Gamma=2t$.

However, the energy of interband-excitations is usually large compared to the superconducting gap. The
collective modes discussed by Leggett\cite{leggett66} are of a fundamentally different nature: If the two
liquids are superfluids, an additional type of tunneling becomes important, {\em i.e.} the simultaneous
tunneling of a {\em pair} of electrons. The tunneling-rate of a pair is usually much smaller than that of a
single electron, and the collective modes which correspond to the dynamical oscillations of pairs between the
two bands have a correspondingly low energy. If the energy is below the gap for pair-excitations, the
dissipation is suppressed, hence these modes can exist by virtue of their low energy. The coupling between the
reservoirs, indicated in Fig. \ref{feynman}b, in this case involves the scattering of a {\em pair} of electrons
in band 1 with momentum (k,-k) to a pair in band 2 with momentum (k',-k') due to the interaction between the
electrons. In Fig. \ref{feynman} this interaction is represented by the exchange of a single boson, but more
complicated processes may be involved. For coupled superconducting bands
$\hbar^{-1}(\nu_s-\nu_d)=\varphi_s-\varphi_d$ is just the phase difference between the two
reservoirs\cite{leggett66}. In this case the pre-factor of the kinetic energy in Eq. \ref{leggetthamiltonian},
$\kappa n_{0}\hbar^2\Gamma^2$, is nothing but the Josephson coupling between the two reservoirs. If the material
is {\em not} a superconductor, {\em a priori} there is no reason for the exciton to be absent. However, the gap
in the superconducting state removes the dissipation for frequencies below the gap, making multi-band
superconductors the best candidates to observe this type of collective mode. At present this type of
relative-phase excitons have been reported in two type of superconductors: (i) the high T$_c$
cuprates\cite{dulic01,grueninger00}, and (ii) MgB$_2$\cite{achterberg02,brinkman04,ponomarev04}. In the
following section we will discuss the former case in detail.

\section{Layered materials}
The cuprates present a rather special case where the interband Coulomb interactions, which were left out of
consideration in the formalism sketched in the previous section, are actually more important than the
compressibility term. Following the Lawrence-Doniach model\cite{lawrence}, we consider each plane as an
individual charge reservoir, or band. Because we will be mainly interested in collective modes involving
currents perpendicular to the planes, we will only deal with the discrete nature of the lattice in the
c-direction. As a result it is convenient to define the Hamiltonian and the equations of motion in terms of
discrete momentum- $\nu_j$ and density-fields $n_j$ (where $j$ is a layer-index) introduced in the previous
section, instead of the continues fields $\nu(r,t)$ and $n(r,t)$, used in section \ref{soundandplasmons}. The
Hamiltonian therefore becomes similar to Eq. \ref{leggetthamiltonian}, but we also want to include a Coulomb
interaction between the layer-charges. We already encountered the Coulomb interaction in section
\ref{soundandplasmons}. Because we want to discuss only propagation perpendicular to the planes, we work in the
limit of zero charge fluctuations parallel to the planes. Integrating over the planes the Coulomb term of Eq.
\ref{lagrangian} becomes equivalent to the interaction between parallel plates of charge. The Coulomb energy of
two positively charged parallel plates decreases linearly as a function of distance. The Hamiltonian density
relevant to the case of c-axis plasmons in layered superconductors is then\cite{marel01}
\begin{eqnarray}\label{layerhamiltonian}
\cal{H}&=& \frac{\kappa n_0^2}{4}
\sum_j \Gamma_{j,j+1}\mbox{}^2 (\nu_j-\nu_{j+1})^2 \nonumber\\
&+&  \frac{1}{2 \kappa n_0^2} \sum_j  n_j^2 -\frac{2\pi e^{*2}}{\epsilon_{\infty}} \sum_{j>l} d_{j,l} n_j n_l
\end{eqnarray}
where $d_{j,l}$ is the distance between the layers with indices $j$ and $l$. The other parameters have already
been discussed in the previous sections. In Ref. \onlinecite{marel96prague,marel01} the discussion has been
limited to the Josephson coupling. In a superconductor $\hbar^{-1}(\nu_j-\nu_{j+1})$ corresponds to the phase
difference $\varphi_j-\varphi_{j+1}$ between neighboring layers. However, superconductivity and long-range phase
coherence are not a pre-requisite for the validity of Eq.\ref{layerhamiltonian}: Under special conditions the
first term of the Hamiltonian corresponds to a Josephson coupling, but more generally it represents the kinetic
energy related to the charge flow. It has a finite value due to the fact that the inertial mass of the charge
carriers is finite.

\section{Plasma dispersion in a single layer material}
We will first discuss the case of a stack of 2D superconducting planes, with lattice constant $d$ along the
c-direction, an compressibility $\kappa$, and an interlayer coupling $\Gamma$. For later use we define here also
the Josephson plasmafrequency, and a dimensionless constant proportional to the incompressibility
\begin{eqnarray}\label{definegamma}
\gamma&\equiv&\frac{\epsilon_{\infty}}{4\kappa n_0^2 d e^{*2}}
\\
\omega_J^2&\equiv&\frac{2\kappa n_0^2 d e^{*2}\Gamma^2}{\epsilon_{\infty}}= \frac{1}{2\gamma}\Gamma^2
\end{eqnarray}
The Hamiltonian equations of motion, $\frac{dn_j(t)}{dt}=\frac{\partial \cal{H}}{\partial\nu_j}$,
$\frac{d\nu_j(t)}{dt} = -\frac{\partial \cal{H}}{\partial n_j} $, give
\begin{eqnarray}\label{singlelayer}
\frac{d^2}{dt^2} n_j&=&\frac{\kappa n_0^2\Gamma^2}{2} \frac{d}{dt} ((\nu_j-\nu_{j+1})+(\nu_j-\nu_{j-1}))
\nonumber\\
&=& \Gamma^2\frac{n_{j+1}+n_{j-1}-2n_j}{2}
\nonumber\\
&+& \omega_J^2 \sum_m\frac{2d_{j,m}-d_{j+1,m}-d_{j-1,m}}{2}n_m
\end{eqnarray}
The distances between the $l$th and the $j$th plane is $d_{j,m}=d|j-m|$. Because
$2|j-m|-|j+1-m|-|j-1-m|=-2\delta_{j,m}$ the only remaining term in the summation over $m$ corresponds to $m=j$.
Substituting the plane wave expression, $n_k=\sum_je^{ikdj}n_j$, in Eq. \ref{singlelayer} we obtain the
frequency-momentum dispersion relation for Josephson-plasmons travelling perpendicular to the planes
\begin{eqnarray}\label{Jdispersion}
\omega(k)=\sqrt{\omega_J^2 + 4\gamma\cos^2(kd/2)}
\end{eqnarray}

For $k=0$ we obtain the usual value for the Josephson plasma frequency. In addition there is an upward
dispersion, which is determined now by the incompressibility parameter $\gamma$, similar to the dispersion in
the continuum model, Eq. \ref{sounddispersion}.

\section{Field effect doping of a single layer material}
\begin{figure}[ht]
\centerline{\includegraphics[width=7cm,clip=true]{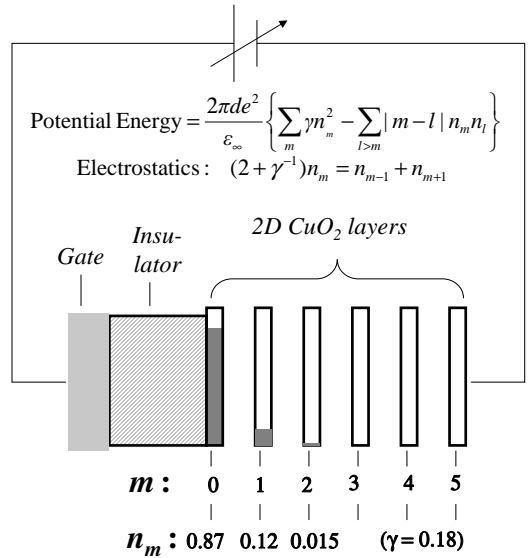}} \caption{Doping profile of a field induced layered
electron gas.} \label{fet}
\end{figure}

Some special consideration deserves the doping of insulating parent compounds by the field effect. Field effect
devices of high Tc materials are a technological challenge. Although this kind of technology has not yet
completely matured, several groups are working into this direction, and a small tuning of the superconducting
transition temperature in a cuprate based field effect device has for example already been realized\cite{ahn03}.

The doping profile in this layered electron liquid model has a simpler form, than the doping profile in a 3D
semiconductor\cite{wehrli01}, and the equations are simpler: All we need to do in order to calculate the doping
profile below the insulating barrier (see Fig\ref{fet}), is to evaluate Eq. \ref{singlelayer} in the static
limit. In other words, we have to equate the lefthand side of the expression to zero. Multiplying both sides of
the expression with $\Gamma^{-2}$ we obtain
\begin{eqnarray}\label{staticsinglelayer}
n_{j+1}+n_{j-1}-\left(2+\frac{1}{\gamma}\right) n_j=0
\end{eqnarray}
which has the following  very simple solution
\begin{eqnarray}\label{solutionstaticsinglelayer}
n_j&=& n_0(1-f)f^j \nonumber\\
f&=&1+ \frac{1}{2\gamma} -\frac{1}{2\gamma}\sqrt{1+4\gamma}
\end{eqnarray}
where $j = 0, 1, 2, ...$. In section \ref{slsco} we will see an example of a cuprate for which $\gamma$ has been
measured experimentally, with the result $\gamma=0.18$. This implies that $f=0.13$. In other words, in a field
effect device for the cuprates (in this example with a lattice spacing of 0.6 nm) the first layer is expected to
receive a fraction $1-f=0.87$ of the total charge induced by the gate of the field effect
device\cite{footnotefet}.



\section{single bi-layer}
Let us now consider a single bi-layer, with a bilayer-distance $d_K$ and a coupling frequency $\Gamma_{1,2}
\equiv \Gamma_{K}$ between layers 1 and 2:
\begin{eqnarray}\label{bilayer}
\frac{d^2}{dt^2} (n_1-n_2)=\kappa n_0^2\Gamma_{K}^2 \frac{d}{dt}
(\nu_1-\nu_2)\nonumber\\
=-\kappa n_0^2\Gamma_{K}^2 \left(\frac{1}{\kappa n_0^2}+\frac{\pi e^{*2}d_K}{\epsilon_{\infty}}
\right)(\nu_1-\nu_2)
\end{eqnarray}
The first term on the righthand side, $\Gamma_K^2$ corresponds to the restoring effect on the non-equilibrium
charge density due to the incompressibility term (the second term in Eq.\ref{layerhamiltonian}), whereas the
second term, $\omega_{K}^2=2\pi e^{*2}d_K \epsilon_{\infty}^{-1}\kappa n_0^2\Gamma_{K}^2$ is due to the
restoring effect of the electric field (the third term in Eq.\ref{layerhamiltonian}). We see, that the resonance
frequency of this mode is
\begin{eqnarray}\label{bilayerfrequency}
\Omega_{K} = \sqrt{\Gamma_{K}^2+\omega_{K}^2} =  \Gamma_{K} \sqrt{1+\frac{2\pi e^{*2}d_{K} \kappa
n_0^2}{\epsilon_{\infty}}}
\end{eqnarray}
The charge and the density appear in the combination $(e^*n_0)^2$, the value of which is independent of whether
one defines $n_0$ and $e^*$ as the density and charge of single electrons, or pairs. In a cuprate superconductor
typically $d \sim 6\cdot 10^{-8}$ cm, $\kappa n_0 \sim 1$eV$^{-1}= 6\cdot 10^{11}$ erg$^{-1}$, the electronic
density is $n_0 \sim 6\cdot 10^{14}$ cm$^{-2}$, and the background dielectric constant $\epsilon_{\infty} \sim
10$. Without the Coulomb interactions, {\em i.e.} assuming that $e^*=0$ in Eq.\ref{bilayerfrequency}, the
frequency of the collective mode is $\Gamma_K$. However, the elementary charge is $e=4.8\cdot 10^{-10}$
erg$^{1/2}$cm$^{1/2}$, and after multiplying all factors we obtain $\sim 3$ for the second term under the square
root. This demonstrates that the 'correction' due to the Coulomb interaction is the dominant contribution to the
internal Josephson resonance frequency of the cuprates. In fact in the first publications on this subject only
the Coulomb term had been taken into account\cite{marel96prague}, whereas the second term of Eq. \ref{bilayer}
(the term considered by Leggett) was neglected. However, for a correct quantitative description of the optical
spectra it is important to take into account all three terms of the hamiltonian Eq. \ref{layerhamiltonian}.

Let us now consider a crystal composed of a stack of bi-layers of the type described above, where the bi-layers
occupy a volume fraction $f$. For example $f$=1/6 in Bi2212. Let us for the moment ignore the inter-bilayer
hopping. With the model above we obtain for the dielectric function\cite{marel01}
\begin{eqnarray}\label{bilayerepsilon}
\epsilon(\omega) = \epsilon_{\infty}\frac{\omega^2-\Omega_{K}^2}
{\omega(\omega+i0^+)-(\Omega_{K}^2-f\omega_{K}^2)}
\end{eqnarray}
This expression predicts an optical absorption at a frequency $(\Omega_{K}^2-f\omega_{K}^2)^{1/2}$, which is
lower than the collective mode of a single bi-layer, $\Omega_{K}$. This reduction of the transverse polarized
collective mode is a consequence of the Coulomb coupling between the bi-layers in the crystal. The longitudinal
mode, {\em i.e.} the frequency for which $\epsilon(\omega)=0$, is at $\Omega_{K}$.

\section{Stack of alternating strong and weak junctions}
\begin{figure}[ht]
\centerline{\includegraphics[width=7cm,clip=true]{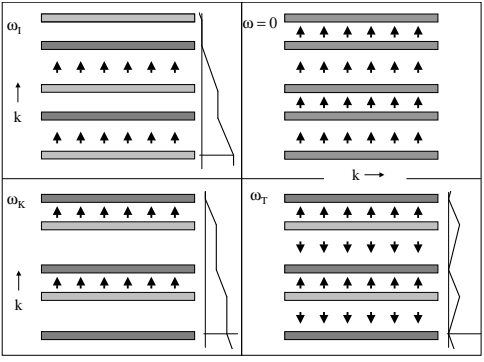}}
 \caption{Snapshot of the currents (arrows) and planar charge fluctuation
 amplitudes (indicated by gray-scales) of the two sets of transverse and
 longitudinal modes with polarization along the c-direction. On the righthand
 side of each plot the voltage distribution is indicated.}
 \label{sketch}
\end{figure}

Because it was assumed that there is no coupling between the bi-layers, the screening at low frequencies is not
contained in Eq. \ref{bilayerepsilon}. In order to describe this effect, Eq. \ref{layerhamiltonian} needs to be
solved with an inter-bilayer coupling taken into account. The result obtained in Refs.
\onlinecite{marel01,bulaevskii02a,bulaevskii02b,helm02a,helm02b,koyama02} is the following
\begin{eqnarray}\label{multilayerepsilon}
\epsilon(\omega) = \frac{\epsilon_{\infty}}{\omega^2}\frac
{(\omega^2-\tilde{\omega}_{+}^2)(\omega^2-\tilde{\omega}_{-}^2)} {\omega(\omega+i0^+)-\tilde{\omega}_{T}^2}
\end{eqnarray}
where the frequencies $\tilde{\omega}_{\pm}$ and $\tilde{\omega}_{T}$ can be expressed in $\Gamma_{K,I}$
(defined in Eq. \ref{layerhamiltonian}) and $\Omega_{K,I}$ (defined in Eq. \ref{bilayerfrequency}). We use the
indices $K$ and $I$ to indicate the bilayer and the inter-bilayer couplings
\begin{eqnarray}\label{multilayerparameters}
\tilde{\omega}_{\pm}^2&=&\frac{1}{2}\left(\Omega_{K}^2+\Omega_{I}^2\right)\pm
\frac{1}{2}\sqrt{\left(\Omega_{K}^2-\Omega_{I}^2\right)^2+4\Gamma_{K}^2\Gamma_{I}^2} \nonumber\\
\tilde{\omega}_{T}^2&=&(1-f)(\Omega_{K}^2+\Gamma_I^2)+f(\Omega_{I}^2+\Gamma_K^2)
\end{eqnarray}
The notation is slightly different from Ref. \onlinecite{marel01}, allowing more transparent expressions. From
this expression we see, that now there are two longitudinal modes (corresponding to $\epsilon(\omega)=0$) and
one transverse mode (a divergence of $\epsilon(\omega)$) at finite frequency. The two extra modes are due to the
out-of-phase oscillation of the inter-plane currents in alternating junctions. This situation has been sketched
in Fig. \ref{sketch}. In the following sections we will discuss some examples, based on materials where this
behaviour has been observed.

\section{Transverse optical plasmon in compounds with weak interlayer tunneling}\label{slsco}
\begin{figure}[ht]
\centerline{\includegraphics[width=7cm,clip=true]{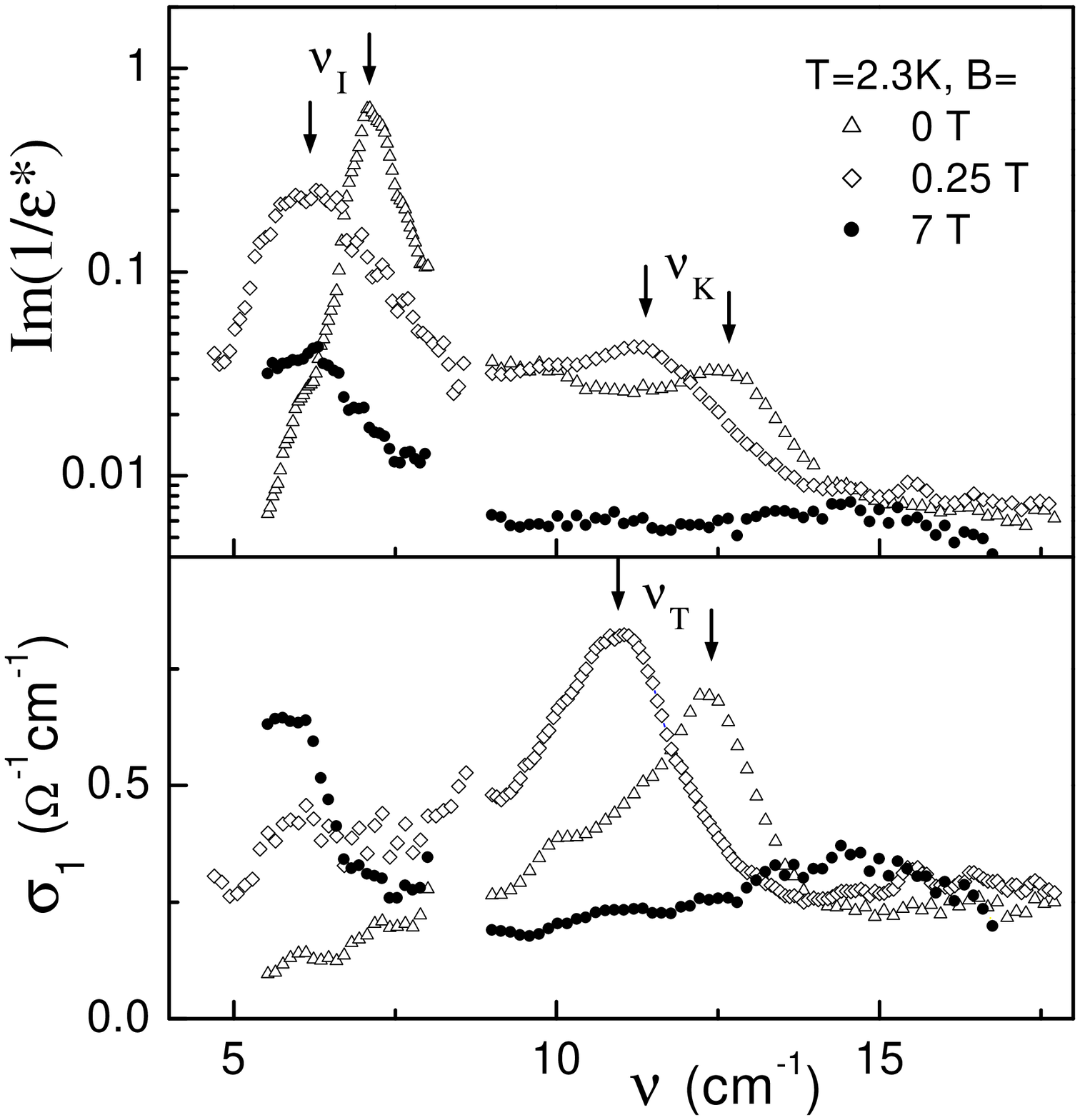}} \caption{Real part of the complex conductivity
$\sigma_1(\omega)$ and loss function $Im(-1/\epsilon(\omega))$ of SmLa$_{0.8}$Sr$_{0.2}$CuO$_{4-\delta}$ along
the c-axis for different magnetic fields. The transverse plasmon $\nu_{T}$ is seen as a peak in $\sigma_1$, the
longitudinal plasmons $\nu_{I,K}$ as peaks in $Im(-1/\epsilon(\omega))$. The data are from Ref.
\onlinecite{pimenov01}.} \label{fcond}
\end{figure}
\begin{figure}[ht]
\centerline{\includegraphics[width=7cm,clip=true]{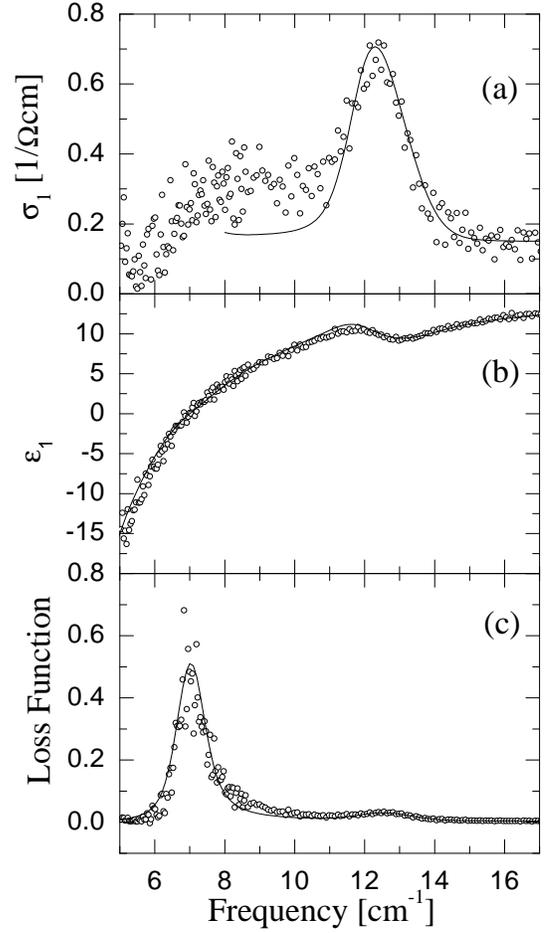}} \caption{ Fit using Eq. \ref{multilayerepsilon},
with the parameters defined in Eqs. \ref{multilayerparameters}, \ref{bilayerfrequency} (solid line) to (a) the
real part of the $c$-axis optical conductivity, (b) Real part of the $c$-axis dielectric function, (c) the loss
function of SmLa$_{0.8}$Sr$_{0.2}$CuO$_{4-\delta}$ at 3K. The open circles are the experimental data. The data
are from Ref. \onlinecite{dulic01}.} \label{fit}
\end{figure}
\begin{figure}[ht]
\centerline{\includegraphics[width=7cm,clip=true]{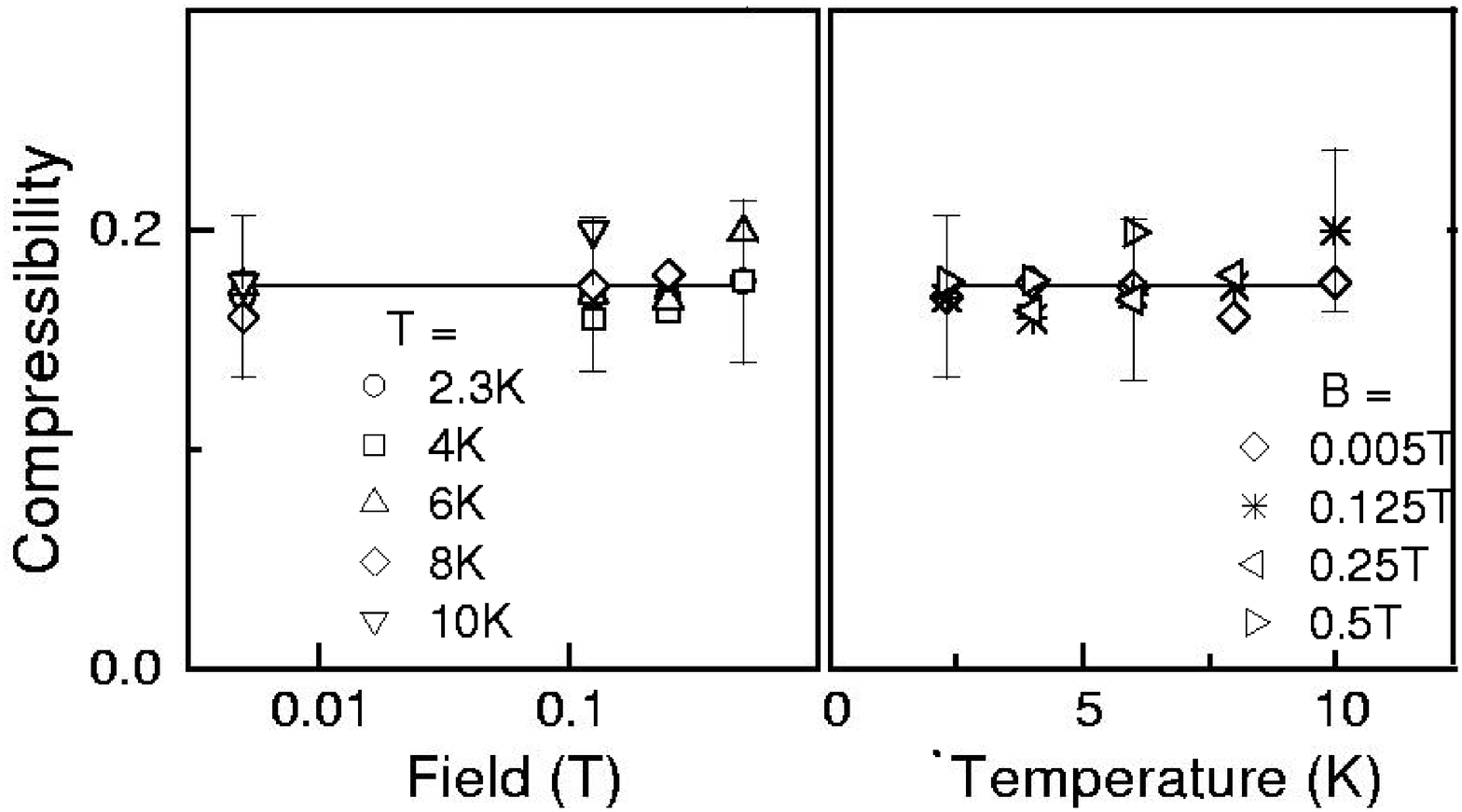}} \caption{Upper panels: magnetic field (left) and
temperature (right) dependence of the electronic incompressibility, $\gamma=\varepsilon_{\infty}/(4\pi
de^{*2}\kappa n_0^{2})$, of SmLa$_{0.8}$Sr$_{0.2}$CuO$_{4-\delta}$. The data are from Ref. \protect
\onlinecite{pimenov01}.} \label{fcomp}
\end{figure}
The existence of {\em two} longitudinal modes and {\em one} associated transverse plasmon mode at finite
frequencies has been confirmed experimentally for the SmLa$_{0.8}$Sr$_{0.2}$CuO$_{4-\delta}$ in a series of
papers\cite{shibata98,shibata01a,shibata01b,kakeshita01,dulic01,shibata02a,shibata02b,pimenov01} (see Figs.
\ref{fig:slsco}, and \ref{fcond}). Measurements of the magnetic field and temperature dependencies of the
longitudinal and transverse plasmons in SmLa$_{0.8}$Sr$ _{0.2}$CuO$_{4-\delta }$ could be successfully described
by the multilayer model explained above, as shown in Fig. \ref{fit}. It is reassuring, that fits to the data
using this model provide for a wide range of temperature and magnetic field the same value for the electronic
incompressibility, $\gamma=\epsilon_{\infty}/(4\pi d e^{*2}\kappa n_0^{2})=0.18$, where $d=1.3$ nm is the
lattice constant, which is twice the spacing between the layers (see Fig. \ref{fcomp}). Note, that the second
factor under the denominator of Eq. \ref{bilayerfrequency} is just $d_K/(2d\gamma)\simeq 1.4$, close to the
estimate given earlier in the discussion. Using the value of $\epsilon_{\infty}=23$ in this compound, and
$a=0.38$nm for the copper-copper distance along the planes, we can use the experimental value of $\gamma$ to
calculate, that $\kappa n_0^2 a^2 = 0.80$ eV$^{-1}$. In a Fermi liquid picture $\kappa n_0^2 a^2$ is exactly the
density of states at $E_F$ per CuO$_2$ unit, $N(0$). For the cuprates $N(0)=$0.8 eV$^{-1}$ is a very reasonable
value.

\section{Transverse optical plasmon and bi-layer splitting in high T$_c$ cuprates}

\begin{figure}[ht]
\centerline{\includegraphics[width=7cm,clip=true]{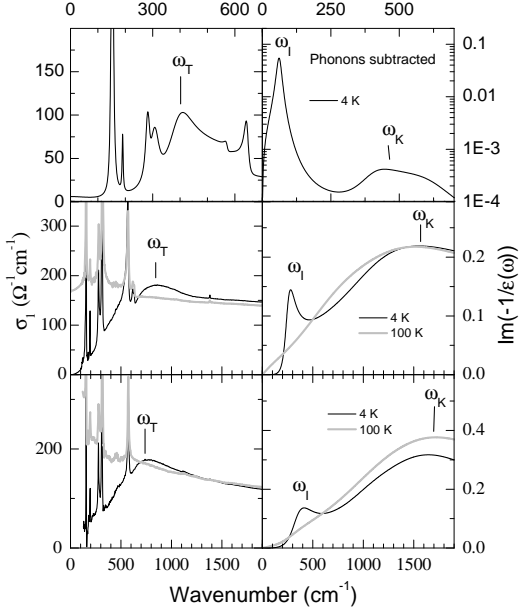}} \caption{C-axis optical conductivity (left) and
energy loss function (right) as a function of wavenumber (in cm$^{-1}$) of under doped (x=6.6, top panels),
optimally doped (x=6.93, middle) and over-doped (x=7.0, bottom panel) YBa$_2$Cu$_3$O$_{7-x}$. The optical
phonons have been subtracted from the loss-functions for clarity. See Refs. \onlinecite{homes95,grueninger00}
for details.} \label{ybcoall}
\end{figure}

The $c$-axis optical conductivity $\sigma_1(\omega)$ of YBCO shows several remarkable
features\cite{homes95,schuetzmann95,tajima97,hauff96,bernhard98}: (1) A very low value compared to band
structure calculations, reflecting the large $\rho_c$. (2) A suppression of spectral weight at low frequencies
already above T$_c$ in underdoped samples referred to as the opening of a ``pseudogap'' (which agrees with the
upturn in $\rho_c$). (3) The appearance of an intriguing broad ``bump'' in the FIR at low T in underdoped
samples. The c-axis optical conductivity of YBCO is one order of magnitude larger than for LSCO near optimal
doping. As a result the relative importance of the optical phonons in the spectra is diminished. C-axis optical
optical conductivity of underdoped \cite{homes95} and optimally and overdoped\cite{grueninger00} YBCO are shown
in Fig. \ref{ybcoall}. Above T$_c$ the optical conductivity is weakly frequency dependent, and does not resemble
a Drude peak. Below T$_c$ the conductivity is depleted for frequencies below 500 cm$^{-1}$, reminiscent of the
opening of a large gap, but not an s-wave gap, since a relatively large conductivity remains in this range.

There is a slight overshoot of the optical conductivity in the region between 500 and 700 cm$^{-1}$, and the
normal state and superconducting state curves cross at 600 cm$^{-1}$. Most of the above mentioned issues can be
clarified by modelling the cuprates, or in particular YBCO, as a stack of coupled CuO$_2$ layers with
alternating weaker and stronger links. Indeed, the transverse mode in the infrared spectrum of optimally and
overdoped YBCO and the above mentioned ``bump'' in underdoped YBCO are well fitted by the multilayer model.
Hence also the ``bump'' in the YBCO c-axis spectra may be regarded as a realization of the ``excitons'' first
considered by Leggett\cite{leggett66}, which involve the relative phase fluctuations of the condensates formed
in two different bands crossing the Fermi
surface\cite{munzar01,munzar03b,boris02,grueninger00,zelezny99,munzar99a,munzar99b,munzar99c,bernhard00}.

This assignment is complicated by the fact, that experimentally the peaks at $\tilde{\omega}_T$ and
$\tilde{\omega}_K$ appear at a temperature higher than the superconducting phase transition. For the underdoped
samples the intensity of these features has been shown to correlate with the intensity of the spin-flip
resonance at $(\pi,\pi)$ seen in neutron spectroscopy\cite{timusk03b}, which is quite far above T$_c$. On the
other hand it appears to be another manifestation of the phenomenon that a strong reduction of dissipation
reveals collective modes, which are otherwise overdamped. In the optimally and overdoped cuprate a wide peak is
visible in the loss-function around 1600 cm$^{-1}$ for all temperatures. The only effect of the transition into
the superconducting state is in this case, that the peak becomes somewhat narrower. This behaviour is
reminiscent of a difficulty which we already encountered in the discussion relating to Figs. \ref{feynman} and
\ref{splitband}: The single particle coupling term $t$ between the layers, Fig. \ref{feynman}a, should in
principle be revealed in the optical spectrum of the normal state as a transition between the bands consisting
of the symmetric and antisymmetric combinations of the two
layers\cite{gartstein94,forsthofer96,atkinson97,dordevic04,shah02}. In the underdoped YBCO this process appears
to be overdamped. On the other hand, the type of process depicted in Fig. \ref{feynman}b can still contribute at
low frequencies, at temperatures where the dissipation becomes sufficiently small. The coherent bi-layer
splitting is gradually restored as we move to the optimally and overdoped region. In this case the two types of
interlayer transport indicated in Fig. \ref{feynman} work in parallel, but there is still a temperature
dependence, like in the other cases. Note that this does not imply nor require the simultaneous presence of both
single electrons and pairs: These are just two different forms of charge transport.

Additional studies of the bi-layer (and tri-layer) materials have provided confirmation of the transverse
optical plasmon in these materials. The spectra of the far-infrared c-axis conductivity exhibit dramatic changes
of some of the phonon peaks, which correlates with the temperature dependence of the transverse optical plasmon.
The most striking of these anomalies can be naturally explained by the local fields acting on the ions arising
from the transverse optical plasmon
oscillations\cite{munzar01,munzar03b,boris02,grueninger00,zelezny99,munzar99a,munzar99b,munzar99c,bernhard00,kovaleva04}.

It is not difficult to extend Eq. \ref{bilayerepsilon} for the dielectric function, to cases where the sequence
along the c-axis is extended to three or more different junctions per period\cite{marel96prague}
\begin{equation}
 \frac{1}{\epsilon(\omega)} = \sum_{m} \frac{z_m\omega^2}
 {\epsilon_{\infty} (\omega^2 - \omega_{J,m}^2)+4\pi i\omega\sigma_m}
 \label{epstot}
\end{equation}
where $z_m$ is the effective volume fraction of the $m$'th junction, and $\sigma_m$ is a parallel dissipative
conductivity of the $m$'th junction. In tri-layer materials such Bi2223 two of the three junctions are identical
because one of the CuO2-layers is a mirror plane. Consequently the expression above has a degeneracy between two
of the three terms in the summation, and the c-axis spectrum should still have one transverse optical plasmon.
Interestingly in this case an additional mode exists where the charge oscillations have even parity around the
mirror-plane\cite{munzar03a}. This electronic mode is observable with Raman spectroscopy\cite{munzar03a}.

If one introduces a single planar defect layer in an otherwise perfectly periodic stack of Josephson coupled
layers, this results in a pronounced satellite line in the real part of the complex resistiviy, whose position
and amplitude depend on the critical current density and on the parameters of the interlayer
coupling\cite{gurevich99}. The extreme narrowness of this plasma-peak could in princple be used to probe the
pairing symmetry using a twist grain boundary configuration.

Random variations of the potential barrier, {\em e.g.} due to chemical disorder, can be taken into account by
replacing the summation over $m$ with a weighted integration over $\omega_{J,m}$. If one assumes for example a
gaussian distribution, a peak appears in the optical conductivity, which coincides with the center-value of the
c-axis plasma frequency\cite{marel96prague}. This effect is present in all published data of the optical
conductivity of LSCO, for example also in Fig. \ref{alexey}. The effect has a strong doping dependence with it's
maximum at exactly $1/8$ doping\cite{dordevic03a}, suggesting an intriguing correlation between disorder in the
interlayer Josephson coupling and the tendency toward stripe-formation. This appears to be a manifestation of a
more general tendency where disorder allows optical absorption by 'forbidden' collective modes. Similar
phenomena have been observed\cite{corson00} along the ab-planes of the cuprates in the THz regime and explained
with a model of a planar disordered array of Josephson junctions\cite{barabash03}.

An alternative way to introduce a more complicated pattern of
Josephson couplings allong the c-axis is obtained by the
application of a magnetic field parallel to the CuO$_2$ planes,
which for YBa2Cu3O6.6 results in inequivalent insulating layers
with and without Josephson vortices. As a result one optical
(transverse) mode appears at around 40 cm$^{-1}$, corresponding to
the antiphase Josephson current oscillations between two
inequivalent junctions\cite{kojima02}.

\section{Summary} In superconductors a rich spectrum of collective
modes can be observed using optical techniques. The simplest case
is where there is one layer per unit cell, for example
La$_{2-x}$Sr$_x$CuO$_4$. Here the Josephson-coupling gives rise to
a single c-axis plasmon. This plasmon is an collective mode of the
phase of the superconducting order parameter. Their coupling to
the electromagnetic field causes a mass-gap of the photons coupled
to the superconductor, providing a small energy scale (and small
budget) demonstration of the Anderson-Higgs mechanism for
generating massive particles. These modes can be described in the
context of the Lawrence-Doniach model. However from the optical
experiments on SmLa$_{0.8}$Sr$_{0.2}$CuO$_{4-\delta}$ we have seen
that it is important to take into account the fact that the
compressibility of the charge-fluid in the layers is finite. The
compressibility term establishes the connection to the model
considered by Leggett for Josephson coupled bands. A directly
observable consequence is the appearance of several additional
collective modes in the optical spectrum, which are related to the
relative phase excitons predicted by Leggett. However, their
energy and optical oscillator strength is strongly affected by the
interlayer Coulomb interaction. These excitons have been observed
for light polarized along the c-axis in a number of cuprate
superconductors. If the number of layers per unit cell is 3 or
more, collective modes of even symmetry appear, which can be
observed with Raman spectroscopy. In addition to chemical
modulation of the interlayer Josephson-coupling, magnetic field
parallel to the planes can result in inequivalent insulating
layers with and without Josephson vortices.
\acknowledgements{This work was supported by the Swiss National Science Foundation through the National Center
of Competence in Research "Materials with Novel Electronic Properties-MaNEP". }


\begin{thebibliography}{100}
\bibitem{footnote1}
The wave-equation for the transverse modes are $
\frac{d^2}{dt^2}\vec{\eta}(r) = \vec{\nabla}\times
\frac{n_0}{\bar{m}} \cdot ( e^*\vec{E}(r) - \vec{\nabla}\int d^3r'
\frac{e^{*2}n(r')}{|r-r'|}- \frac{\vec{\nabla}n(r)} {\kappa n_0^2}
- \frac{\vec{\nabla}\nu(r)}{\tau})$. For waves travelling along
one of the principle axis of the mass tensor:
$\frac{d^2}{dt^2}\vec{\eta}(r) = -\frac{e^*n_0}{m}
\frac{d}{dt}\vec{B}(r,t)$. Solving this together with the Maxwell
equations yields the polariton dispersion relation
$k^2c^2=\omega^2 \epsilon(\omega)$.
\end{thebibliography}
\end{document}